\pdfoutput=1
\RequirePackage[T1]{fontenc}
\RequirePackage{fix-cm}
\RequirePackage{lmodern}

\documentclass[letterpaper,12pt,pdftex,unicode]{article}
\usepackage{stix2}
\newtheorem{proposition}{Proposition}
\newcommand{\ket}[1]{|#1\rangle}

\begin{document}
\title{Breeding protocols are advantageous for finite-length entanglement distillation}
\author{Ryutaroh Matsumoto}
\date{February 1, 2024}

\maketitle
\begin{abstract}
  Bennett et al.\ \cite{bennett96} proposed a family of protocols for
  entanglement distillation, namely, hashing, recurrence and breeding protocols.
  The last one is inferior to the hashing protocol in the asymptotic regime
  and has been investigated little.
  In this paper, we propose a framework of converting a stabilizer
  quantum error-correcting code to a breeding protocol,
  which is a generalization of the previous conversion methods by
  Luo-Devetak and Wilde.
  Then, show an example of a stabilizer that gives a breeding protocol
  better than hashing protocols, in which the finite number of maximally
  entangled pairs are distilled from the finite number of
  partially entangled pairs.
\end{abstract}

\section{Introduction}
The quantum internet has attracted much attention recently \cite{azuma15,cacciapuoti20,illiano22}.
The entanglement distillation (or purification)
protocols (EDPs) by Bennett et~al.\ \cite{bennett96}
is a fundamental theoretical component for the quantum internet.
An EDP distills the maximally entangled states from noisy pairs
of entanglements between two parties, usually called Alice and Bob.
In the early stage of EDP research,
protocols were found by ad-hoc manners \cite{bennett96,deutsch96,deutsch98}.
Later, systematic constructions \cite{shor00,chuangnielsen,matsumoto-epp}
of EDPs from the quantum error-correcting codes were proposed.
Among those systematic constructions \cite{shor00,chuangnielsen,matsumoto-epp},
the quantum privacy amplification (QPA)
protocol \cite{deutsch96,deutsch98} can only be constructed
by \cite{matsumoto-epp}, which includes \cite{shor00,chuangnielsen}
as its special cases.

In the newer construction \cite{matsumoto-epp},
Alice uses the complex conjugation of Bob's local operation (LO) as her LO,
while the previous constructions \cite{shor00,chuangnielsen} use the
same LO by both Alice and Bob.
The necessity of the complex conjugation was stated earlier by
Horodecki and Horodecki \cite{horodecki99}.
Subsequently,
improved EDPs are constructed \cite{watanabe06a,meikoudai}
based on \cite{matsumoto-epp}
with the two-way classical communications,
with which the complex conjugation was indispensable.

On the other hand,
Bennett et~al.\ proposed three kinds of the EDPs,
the hashing protocol, the recurrence protocol and the breeding protocol.
The breeding protocol assumes the preshared maximally entangled states,
while the other two protocols have no use of preshared maximally entangled states.
The recurrence protocol uses two-way classical communication,
while the other two protocols do not.
Later, Duo and Devetak \cite{luo07} and Wilde \cite[Section 9.2]{wildephd}
proposed a conversion method
of binary stabilizers to EDPs, which did not take the complex conjugation \cite{horodecki99}
into their considerations. Consequently,
their conversion method cannot handle encoding operators of quantum messages
to quantum stabilizer codewords when they cannot be written as a real matrix.
The breeding protocol is still studied recently, e.g.\ \cite{dur21},
but the research of breeding protocol does not seem very active.
The reason of inactivity is probably that there always exists an asymptotically
better or equally good hashing protocol than any given breeding protocol
\cite{bennett96}.
However, an EDP always distills 
the finite number of maxinally
  entangled pairs from the finite number of
  partially entangled pairs.
  The possibility of constructing a recurrence protocol better than any hashing protocols
  has not been investigated throughly.

The purposes of this paper are two-fold.
Firstly, we add the complex conjugation to  the previous conversion method
\cite{luo07,wildephd}, which enables conversion of any (binary and non-binary) stabilizer
with any encoding operators.
The encoding operators of non-binary stabilizer codes usually cannot be written
as real matrices.
Secondly,
we show examples of breeding protocols better than
any hashing protocol constructed from stabilizers,
distilling maximally entangled pairs from the same number of
partially entangled pairs.

\section{Preliminaries}
\subsection{Stabilizer codes}
In this paper,
we consider a $p$-dimensional quantum system $\mathcal{H}_p$
as a unit of quantum information processing,
where $p$ is a prime number.
Let $\mathbf{F}_p$ be the finite field with $p$ elements.
For two vectors $(\vec{a}|\vec{b})$, $(\vec{c}|\vec{d}) \in \mathbf{F}_p^{2n}$,
their symplectic inner product is defined by
\[
\langle \vec{a}, \vec{d}\rangle_E - \langle \vec{b}, \vec{c}\rangle_E,
\]
where $\langle \cdot$, $\cdot\rangle_E$ denotes the standard
Euclidean inner product.
For a vector $(\vec{a}|\vec{b}) \in \mathbf{F}_p^{2n}$,
its symplectic weight $w_s(\vec{a}|\vec{b})$ is defined by
$|\{ i : (a_i, b_i) \neq (0,0) \}|$, where $a_i$ (resp.\ $b_i$)
is the $i$-th component of $\vec{a}$ (resp.\ $\vec{b}$).
It is now well-known that a quantum stabilizer code can be described by
an $\mathbf{F}_p$-linear space $C \subset C^{\perp s} \subset
\mathbf{F}_p^{2n}$.
The code $C$ encodes $k$ qudits in $\mathcal{H}_p^{\otimes k}$
into $n$ qudits in $\mathcal{H}_p^{\otimes n}$,
where $k = n - \dim C$,
and it can correct $\lfloor (d-1)/2 \rfloor$ errors,
where $d = \min \{ w_s(\vec{a}|\vec{b})$ : 
$(\vec{a}|\vec{b}) \in C^{\perp s} \setminus C \}$.
When we have $d = \min \{ w_s(\vec{a}|\vec{b})$ : 
$(\vec{a}|\vec{b}) \in C^{\perp s} \setminus C \} =
\min \{ w_s(\vec{a}|\vec{b})$ : 
$(\vec{a}|\vec{b}) \in C^{\perp s} \setminus \{\vec{0} \}$,
the code $C$ is said to be \textbf{pure}.
Besides errors,
$C$ can also corrects $d-1$ erasures, where an erasure is
an error whose position is available to a decoder
among $n$ qudits in a codeword \cite{grassl97,macwilliams77}.
Simultaneously, $C$ can corrects $t$ errors and $e$ erasures
provided that $2t + e < d$.
The quantum code $C$ is said to be an $[[n,k,d]]_p$ stabilizer code.

Let $\{ \vec{h}_1$, \ldots, $\vec{h}_{n-k}\}$
be a basis of $C$.
In the standard decoding procedure with $C$,
a decoder measures observables corresponding to $\vec{h}_i$
for $i=1$, \ldots, $n-k$,
and obtains syndromes $s_i \in \mathbf{F}_p$ for $i=1$, \ldots, $n-k$.
Then the decoder decides a suitable unitary operator $U$ according
to $s_1$, \ldots, $s_{n-k}$.
and applies $U^{-1}$ to the measured quantum state.

\subsection{Entanglement distillation protocols and stabilizer codes}
Let $\{ \ket{i} : i \in \mathbf{F}_p \}$ be an orthonormal basis of $\mathcal{H}_p$,
and define a maximally entangled state
\[
\ket{\Psi_p} = \frac{1}{\sqrt{p}} \sum_{i \in \mathbf{F}_p} \ket{i} \otimes \ket{i}
\]
in $\mathcal{H}_p \otimes \mathcal{H}_p$.
Let $\rho$ be a partially entangled state in $\mathcal{H}_p \otimes \mathcal{H}_p$.
We assume that $\rho^{\otimes n}$ is shared between two parties,
called Alice and Bob.
The purpose of an entanglement distillation protocol (EDP)
is distillation of a nearly maximallly entangled state close to $\ket{\Psi_p}^{\otimes k}$
from $\rho^{\otimes n}$ by local operations and classical communications (LOCC)
between Alice and Bob.

It is known \cite{shor00,chuangnielsen,matsumoto-epp}
that an $[[n,k,d]]_p$ stabilizer $C \subset \mathbf{F}_p^{2n}$ can be converted to
an EDP as follows.
For $\vec{v} = (\vec{a}|\vec{b}) \in \mathbf{F}_p^{2n}$,
let $\vec{v}^\star = (\vec{a}|-\vec{b}) \in \mathbf{F}_p^{2n}$.
When $\{ \vec{h}_1$, \ldots, $\vec{h}_{n-k}\}$
be a basis of $C$,
we denote by $C^\star$ the linear space spanned by
$\{ \vec{h}_1^\star$, \ldots, $\vec{h}_{n-k}^\star\}$.
Note that $C \subset C^{\perp s}$ implies $C^\star \subset (C^\star)^{\perp s}$.
EDP constructed from $C$ can be executed as follows \cite{matsumoto-epp}.
Note that the following protocol is a generalization of \cite{shor00,chuangnielsen}.
The generalization allows the QPA protocol \cite{deutsch96,deutsch98} to be interpreted
as its special case.

The following protocol distills $k$ pairs of maximally entangled state
$\ket{\Psi_p}^{\otimes k}$ from
$n$ paris of noisy entangled state $\rho^{\otimes n}$
in $(\mathcal{H}_p \otimes \mathcal{H}_p)^{\otimes n}$,
provided that in $\rho^{\otimes n}$
there are at most $t$ errors and $e$ erasures with $2e + t <d$.

\begin{enumerate}
\item Alice performs the measurement corresponding to the stabilizer code
  $C^\star$. The syndrome is denoted by $a_1$, \ldots, $a_{n-k} \in \mathbf{F}_p$.
\item Bob performs the measurement corresponding to the stabilizer code
  $C$. The syndrome is denoted by $b_1$, \ldots, $b_{n-k} \in \mathbf{F}_p$.
\item Alice sends the  syndrome $a_1$, \ldots, $a_{n-k}$ to Bob.
\item Bob performs a decoding process regarding $a_1+b_1$, \ldots, $a_{n-k}+b_{n-k}$
  as the ordinary syndrome in the standard decoding procedure with $C$.
  (Note that the sign of $a_i$ is negated from \cite{matsumoto-epp}.)
\item Alice and Bob apply the inverse of encoding operators of the quantum stabilizer codes
  $C^\star$ and $C$.
\item Alice and Bob retain $k$ information qudits and discard the rest $n-k$ qudits,
  regarding their $n$ qudits as codewords of stabilizer codes.
\item (Only when the two-way communications are available.)
  When $a_1+b_1$, \ldots, $a_{n-k}+b_{n-k}$ indicate low fidelity,
  Alice and Bob discard the resulted quantum state, as done in the recurrence protocol \cite{bennett96}.
  \end{enumerate}

\subsection{Entanglement-assisted stabilizer codes}
As mentioned above,
the construction of stabilizer codes requires self-orthogonal linear spaces,
which imposes additional difficulty.
Brun et al.\ \cite{brun06} proposed entanglement-assisted stabilizer codes
to remove the requirement of self-orthogonality,
and enabled us to construct a quantum error-correcting code
from any linear space over a finite field.
The following formulation is from \cite{galindo19,galindo19e}.

Let $D \subset \mathbf{F}_p^n$ with $n-k = \dim D$,
$c = \dim D - \dim D \cap D^{\perp s}$, and
$d = \min \{ w_s(\vec{a}|\vec{b}) : D^{\perp s} \setminus D \}$.
Under the assumption that an encoder and a decoder share
$c$ maximally entangled state $\ket{\Psi_p}^{\otimes c}$,
the encoder encodes $k+c$ qudits into a codeword of
$n$ qudits.
There always exists an $[[n+c, k+c]]_p$ stabilizer code $C$ whose
punctuation is equal to $D$ (see \cite{galindo19,galindo19e}).
Let $\vec{h}_1$, \ldots, $\vec{h}_k$ be a basis for $C$.
The decoder measures observables corresponding to
$\vec{h}_1$, \ldots, $\vec{h}_k$ of the received quantum state  and
$c$ halves of $\ket{\Psi_p}^{\otimes c}$
possessed by the decoder.
Then $t$ errors and $e$ erasures in the codeword can be corrected
by the decoder if $2t+e < d$.
This coding method is called an $[[n, k+c, d; c]]_p$ entanglement-assisted
quantum error-correcting code (EAQECC).

\begin{proposition}\label{prop1} \cite{ueno22}
An $[[n-c, k-c, d; c]]_p$ EAQECC
can be constructed from an $[[n, k, d]]_p$ \textbf{pure}
stabilizer code $D$
\end{proposition}

\section{Proposed EDPs from EAQECCs}
In this section, we propose a construction procedure of an EDP from an
$[[n,k,d;c]]_p$ EAQECC $D$.
Also assume that $D$ is a punctuation of an $[[n+c, k]]_p$ stabilizer code $C$
whose basis is $\vec{h}_1$, \ldots, $\vec{h}_k$.

As the breeding protocol in \cite{bennett96},
Alice and Bob share $c$ maximally entangled state $\ket{\Psi_p}^{\otimes c}$.
Their goal is to distill $k$ pairs of maximally entangled state 
$\ket{\Psi_p}^{\otimes k}$ from the $c$ shared
maximally entangled state $\ket{\Psi_p}^{\otimes c}$
and $n$ pairs of noisy entangled state $\rho^{\otimes n}$
in $(\mathcal{H}_p \otimes \mathcal{H}_p)^{\otimes n}$.
It is executed as follows:

\begin{enumerate}
\item Alice performs the measurement corresponding to the stabilizer code
  $C^\star$ for $n$ halves of the noisy entangled state and
  $c$ halves of $\ket{\Psi_p}^{\otimes c}$.
  The syndrome is denoted by $a_1$, \ldots, $a_{n-k} \in \mathbf{F}_p$.
\item Bob performs the measurement corresponding to the stabilizer code
  $C$
   for $n$ halves of the noisy entangled state and
   $c$ halves of $\ket{\Psi_p}^{\otimes c}$.
   The syndrome is denoted by $b_1$, \ldots, $b_{n-k} \in \mathbf{F}_p$.
\item Alice sends the  syndrome $a_1$, \ldots, $a_{n-k}$ to Bob.
\item Bob performs a decoding process regarding $a_1+b_1$, \ldots, $a_{n-k}+b_{n-k}$
  as the ordinary syndrome in the standard decoding procedure with $C$.
  (Note that the sign of $a_i$ is negated from \cite{matsumoto-epp}.)
\item Alice and Bob apply the inverse of encoding operators of the quantum stabilizer codes
  $C^\star$ and $C$.
\item Alice and Bob retain $k$ information qudits and discard the rest $n+c-k$ qudits.
\item  (Only when the two-way communications are available.)
  When $a_1+b_1$, \ldots, $a_{n-k}+b_{n-k}$ indicate low fidelity,
  discard the resulted quantum state, as done in the recurrence protocol \cite{bennett96}.
  
  \end{enumerate}

By the almost same argument as \cite{matsumoto-epp},
the above protocol can correct $t$ errors and $e$ erasures if $2t+e < d$.
It produces $k$ pairs of maximally entangled state $\ket{\Psi}_p$,
while destroying $c$ pairs.
The net production of $\ket{\Psi}_p$ is $k-c$.

\section{Examples}
The following knowledge on existences of code parameters are
obtained from \cite{codetable}.
By Proposition \ref{prop1},
an $[[n,k,d]]$ \textbf{pure} stabilizer code,
can be converted to $[[n-c, k, d, c]]_p$ EAQECC if $c < d$.

There exists an $[[6,4,2]]$ pure stabilizer code, which can be converted
to an EDP distilling $3$ pairs of $\ket{\Psi}_p$ from $5$ noisy
pairs of entangled state while correcting $1$ erasures.
On the other hand,
there is no $[[5,3,2]]_2$ stabilizer code, and
our previous proposal \cite{matsumoto-epp} without the two-way communications
cannot produce an EDP with the same performance.

Similarly, the existence and the non-existence of code parameters
in Table \ref{tab1} show that our proposal in this paper
can construct EDPs exceeding the ones from the previous construction procedure \cite{matsumoto-epp}.

\begin{table}
  \caption{Existence and non-existence of binary code parameters \protect\cite{codetable}}\label{tab1}
  \centering\begin{tabular}{cc}
    exist & not exist\\\hline
    $[[6,4,2]]$ & $[[5,3,1]]$\\
    $[[8,3,3]]$ &  $[[7,2,3]]$\\
    $[[10,2,4]]$ & $[[9,1,3]]$\\
    $[[16,2,6]]$ & $[[15,1,6]]$\\
    $[[21,15,3]]$ & $[[20,14,3]]$
  \end{tabular}
\end{table}

\section{Conclusion}
In this paper, we generalized previous methods of EDP constructions from
stabilizer codes \cite{shor00,chuangnielsen,matsumoto-epp,luo07,wildephd},
and enabled conversion of any $p$-adic quantum stabilizers and their encoding
operators to breeding EDPs. Then, we explicitly provided examples of less-investigated
breeding protocols, which have performances better than any hasing protocols
constructed from quantum stabilizers.

We have not considered possitility of two-way classical communications (2-CC)
It is known that adding 2-CC to the hashing protocols provides much better
performace \cite{bennett96,watanabe06a,meikoudai}.
The investigation of 2-CC with breeding protocols is an important future research agenda.

\section*{Acknowledgment}
The author would like to thank Prof.\ Mark W. Wilde for
informing me of \cite{luo07,wildephd}, which
corrected overstating of the novelty in an earlier manuscript.
This research is in part supported by the Japan Society for the Promotion of Science
under Grant No.\ 23K10980.


\end{document}